\documentstyle[pre,aps,multicol,epsf]{revtex}
\begin{document}
\widetext

\draft
\title{
Identity of the universal repulsive-core singularity  \\
  with  Yang-Lee edge criticality }
\author{Youngah Park$^{1,2}$ and Michael E. Fisher$^{1}$ }
\date{July 1999}
\address
{
$^1$ Institute of Physical Science and  Technology,
University of Maryland. College Park, MD 20742 \\
$^2$  Department of Physics, Myong Ji University,
 Yongin, Kyonggi-do,  449-728, Korea 
   } 
\maketitle
\thispagestyle{empty}
\begin{abstract}
\hspace{1em} Lattice and continuum fluid models with repulsive-core interactions 
typically display a dominant, critical-type 
singularity on the real, {\it  negative} activity axis.
Lai and Fisher  recently suggested, mainly on numerical grounds,
that this repulsive-core singularity is  universal and in the same class
as the Yang-Lee edge singularities, which 
arise above criticality at  
 complex activities  with {\it positive}  real part.
A general analytic demonstration of this 
 identification is presented here  using 
a field-theory approach with separate representation of the repulsive 
and attractive parts of the  pair interactions.

\end{abstract}
\begin{multicols}{2}

\section {Introduction}
In 1952 Lee and Yang \cite{leeyang} proved  that
the zeroes of the partition function of a
ferromagnetic Ising model (or equivalently, a lattice gas with single-site 
hard cores) as a function of  complex magnetic
field $H$ (or chemical potential $\mu$)  are confined to the imaginary $H$
axis for  real
temperatures $T$.
Yang and Lee argued  \cite{yanglee} that for a system above 
its critical temperature
$T_c$, the partition function    must be nonzero 
throughout some neighborhood of the real axis 
in the complex reduced magnetic-field plane,
\begin{equation}
h = H/{k_B T } \equiv  h^{\prime} + i h^{\prime \prime}.
\end{equation}
Thus for $T> T_c$, a gap   free of zeros will be found on the imaginary 
$h$ axis with edges at,
say,  $ \pm i h_{\sigma}  (T)$. 
Equivalently, in fluid-language, for $T >T_c$, there 
will be a gap in the complex activity plane with
edges at 
\begin{equation}
z = z_{\sigma}^{\prime} (T) \pm i z_{\sigma}^{\prime \prime} (T) , 
\end{equation}
as illustrated in Fig. 1. 
Here we define the reduced activity in $d$ spatial dimensions by 
\begin{equation}
z = e^{ \beta \mu} v_0 / {\Lambda}_T^d , 
\end{equation}
where $\beta = 1/k_B T$, while $v_0$ is a microscopic reference 
volume (taken as the cell volume for a lattice gas)  and ${\Lambda}_T $ 
is the thermal de Broglie wavelength.

If one  defines the  density of Yang-Lee zeros, $g(h^{\prime \prime}) $, 
so that  when 
$N$, the number of spins (or lattice sites)   becomes infinite,  
$N g( h^{\prime \prime} ) d h^{\prime \prime} $ approaches the number 
of zeros between $i h^{\prime \prime} $ and
$ i ( h^{\prime \prime} + d  h^{\prime \prime}) $ on the imaginary $h$ axis, 
one must have 
$ g (  h^{\prime \prime} ) \equiv  0  \hspace{1em}   {\rm for}
\hspace{1em}
 | h^{\prime \prime} |
< h_{\sigma} (T),\hspace{1em}  {\rm for} \hspace{1em}  T>T_c $ .
Kortman and Griffiths \cite{Kort:Grif} pointed out that the density  
of zeros beyond the gap should be  expected to exhibit a power law
singularity---the Yang-Lee edge singularity \cite {MEF}---of the form 
\begin{equation}
g (  h^{\prime \prime} ) \sim  | | h^{\prime \prime} | - 
h_{\sigma}  (T) |^{\sigma}
\hspace{1em}   {\rm for} \hspace{1em}  | h^{\prime \prime} |
\rightarrow  h_{\sigma}  (T)- .
\end{equation}
This singularity proves analogous to an 
ordinary critical point obeying scaling laws and exponent relations 
although there is only one relevant scaling field [3, 4].
Thus the basic exponent $\sigma$ in (4) satisfies 
$ \sigma = 1/{\delta} = ( d - 2 + \eta ) / (d + 2 - \eta )  $\cite{MEF}.

 In a  general renormalization group analysis,
it was shown \cite{MEF} that the  field theory controlling 
the Yang-Lee fixed point is described by  
a pure imaginary $ i  {\varphi}^3 $ coupling. 
This leads to a critical dimension $d_c = 6$ above which 
the classical, mean-field value \cite{MEF} $ \sigma = {\textstyle \frac{1}{2}} $ 
applies.
To first order in  $ \epsilon = 6 -d >0  $ one has
$ \sigma (d) = {\small \frac{1}{2}}  - {\small \frac{1}{12}} \epsilon $ and 
$ \eta = - {\small \frac{1}{9}} \epsilon $. 
Expansions of $\sigma (d)$ to  order  
${\epsilon}^3 $  are known \cite{Mckane} and one has  
$\sigma (1) = - {\small \frac{1}{2}} $ \cite{MEF2}, 
$ \sigma (2) =  - {\small \frac{1}{6} } $,  and finds, numerically,
$\sigma (3) \simeq   0.088 $ \cite{lai} : 
see Lai and Fisher \cite{lai} who review 
previously known relationships of the Yang-Lee edge singularity to a number of 
different problems,   
specifically, isotropic branched polymers or,  equivalently, 
undirected lattice animals with or without loops allowed, 
in $(d+2)$ dimensions [8, 9];  
Anderson localization 
\cite{lubmck}; and {\it directed} branched polymers (or directed, loop-free lattice animals) in $d+1$ dimensions \cite{cardy}.


By contrast, an apparently quite different type of singularity arises in fluid systems
when the particle  interactions have repulsive cores. If the pair interaction potentials
are purely repulsive (i.e., positive),
the (reduced) 
cluster integrals, $b_n (T) $, in the activity  or fugacity series for the (reduced) pressure,
namely,
\begin{equation} 
{ \bar p } \equiv  \frac{ p v_0 }{ k_B T } =
\sum_{ n =1}^{ \infty} b_n (T)  z^n  , 
\end{equation} 
are known \cite{gro} to alternate in sign: this   implies  a dominant singularity 
on the 
{\it negative } $z$-axis that determines the radius of convergence, $R$, of the 
series: see Fig. 1.  
In the vicinity of this  singularity, say  at $ z = z_0(T)  = -R$, the reduced 
pressure can be written as
\begin{eqnarray}
{\bar p } (z)&  =&  {\bar p}_0 + {\bar p}_1 (z - z_0 ) +
 {\bar p}_2 ( z - z_0 )^2 +  \cdots \\ \nonumber
 &             & +   \hspace{1em}  P (z - z_0 )^{\phi} [1 + a_{\theta} ( z - z _0 )^{\theta} + 
\cdots ] + \cdots  ,
\end{eqnarray} 
where the exponents    $\phi$ and $\theta$ are anticipated 
to be  nonintegral.

Indeed, in 1984,  Poland \cite{poland} studied a variety of lattice models 
and a continuum 
fluid of  hard squares and proposed that this repulsive-core
singularity is characterized by a {\it universal} exponent $\phi (d) $.
Subsequent confirmation came from   Baram and Luban \cite{baram} who 
investigated further models including
dimers on lattices, parallel hypercubes in
continuum space and the {\it soft}-core single-component Gaussian-molecule model.
More recently, Lai and Fisher \cite{lai} found similar behavior 
in a {\it binary} 
Gaussian-molecule mixture using very long series expansions for $d = 1, \cdots , 6$.
In that case the singularity at $z=z_0$ was  drawn out into a continuous locus;
but the estimates for $ \phi (d) $ supported the universality 
hypothesis. (Precise estimates, 
for the leading ``correction-to-scaling" exponent
$\theta (d)$, in (6) for all $d$, 
including $ \theta (1) = 1$, $ \theta (2) = {\textstyle \frac{5}{6} } $,
and
$ \theta (3) \simeq 0.62$, were also generated \cite{lai}.)

However,  Lai and Fisher \cite{lai} noticed in particular that, when compared 
with previous knowledge
about the Yang-Lee edge exponent $ \sigma  (d) $, the exact results 
for $    \phi  (d) $ for $d =1$ and $d = \infty $ \cite{hauge}  and 
for $ d = 2 $
\cite{baram,baxter}, and the various numerical estimates 
\cite{lai} for $d \ge 2$,
strongly suggested the identification
\begin{equation}
\phi (d) = \sigma (d) + 1 \hspace{3em} {\rm for } \hspace{1em}  {\rm all } 
\hspace{1em} d .
\end{equation}
Hence they proposed \cite{lai} that the 
{\it universal repulsive-core singularity belongs to the Yang-Lee edge 
critical universality  class}.

In analytical support of this identification 
(for general $d$) they appealed to earlier work by
Kurtze and Fisher \cite{kurtze} who had proved that when 
$T \rightarrow \infty $ the Yang-Lee edge
singularity in ferromagnetic Ising models 
precisely describes the dominant singularity on the 
negative $z$ axis of a fluid of {\it hard dimers} 
on the same lattice, each dimer 
occupying one bond and two adjacent lattice sites.
This correspondence with a dimer gas was carried further by 
Shapir \cite{shapir} who used field-theoretic arguments 
to demonstrate the identity generically for all $T > T_c$. 
However, dimers are rather special objects with orientational 
degrees of freedom (although these seem, {\it post facto},
to have no effect on $ \phi (d) $ \cite{baram}).
Furthermore, Shapir's field-theoretic approach was rather special 
and did not seem extendable to
more general particles with 
soft repulsive cores, with the additional presence of 
attractive forces, or to systems also displaying critical behavior 
for positive $z$ \cite{lai} (as illustrated in Fig. 1).
 
In this article we repair this gap in the theory.
Specifically, we consider a general single-component 
fluid with a pair interaction potential,
$U ( {\bf r}  ) $, which contains both repulsive {\it and} attractive parts:
of course, repulsive terms are always essential to ensure 
thermodynamic stability.
To be concrete and explicit we analyze lattice systems in 
which multiple occupancy of a site is forbidden; however, it must be stressed
that the repulsive interactions we consider are {\it not} confined to 
such trivial single-site 
hard cores.  
On the contrary, essentially we suppose only  
that the potential $U ( {\bf r} )$ is of finite range and may be 
decomposed according to 
\begin{equation}
 {\hat U}  ( {\bf  k }) = 
\sum_{ {\bf R } \neq 0 } e^{ i {\bf k } \cdot { \bf R } } 
U ( {\bf  R } ) = { \hat w } ( {\bf k } ) - { \hat v } ( {\bf k } ) ,
\end{equation}  
where the sum runs over lattice sites 
$ {\bf R} = {\bf  r }_j  $ while the repulsive and attractive parts, 
${ \hat w } ( {\bf  k } )$  and $  { \hat v } ( {\bf k } )$, respectively, are both 
positive (if they do not vanish identically). 
More specifically 
we will use
\begin{equation}
 {\hat w ({\bf k} ) } = {\hat w}_0 ( 1  - k^2 a^2    + \cdots  )  > 0 ,  
\end{equation}
\begin{equation}
{\hat v  ({\bf k} ) } = {\hat v }_0 ( 1-   k^2  b^2   + \cdots  ) >  0 ,
\end{equation}  
so that $a$ and $b$ represent the interaction ranges of the repulsive 
and attractive  components. 
The real-space potentials acting between sites $j$ and $k$ ($j \neq k$) , namely, 
$ w_{jk} = w ( {\bf  r}_k - {\bf r}_j ) $ and  
$ v_{jk} = v ( {\bf r}_k - {\bf r}_j ) $, follow
by Fourier inversion.
For $d=3$ one may imagine Yukawa forms:
$ w ( {\bf r}) = W_0 e^{ - r/a} /r $ and 
$ v( {\bf r}) = V_0 e^{ - r/b } /r $ with $ a< b$; but that is certainly not 
essential. Likewise, the leading lattice  isotropy assumed for convenience 
in (9) and (10) is not necessary.
Our treatment extends 
straightforwardly to multicomponent fluids \cite{lai} and, at least formally, 
generalizes readily to continuum systems.

On the basis of (8)-(10), we develop a field-theoretic analysis and show that there is,
in general, a repulsive-core singularity at some $z_0 (T) $ on the 
negative activity axis (see Fig. 1) that is described 
within an LGW renormalization group  framework by 
a fixed point Hamiltonian with a purely 
imaginary cubic coupling, $i { \varphi}^3$, and, hence, lies in the same universality class as 
Yang-Lee edge criticality.
A more or less novel aspect of our treatment is that 
{\it separate}---sine-Gordan \cite{jaffe} and Kac-Hubbard-Stratonovich (KHS) 
\cite{kac,hubbard,stra}---transformations are used for handling 
the repulsive and attractive parts of the interaction potential:
compare with Refs. 23 and 24.

\section{FORMULATION}

For a lattice of volume $V$ with sites labeled 
$j,k,l=1,2,{\cdots}$,    $ {\cal N} = V/v_0$, let
${\rho}_j =0$ or $1$ according 
as site $j$ is or is not occupied by a particle.
Then, recalling (3) and (8)-(10) {\it et seq}., the grand partition
function for the lattice system is

\begin{equation}
{\cal Z} (T,z) =  {\rm Tr}_{ \rho  }^{\tiny  {\cal N} }  \left\{  
z^{ {\textstyle { \sum_j  }   } \hspace{0.1em} {\rho}_j }
\exp[ - {\textstyle \frac{1}{2} } \beta  \sum_{j \neq k}    
    \rho_j    
( w_{jk}  - v_{jk} )  \rho_k ]  
\right\} .
\end{equation}
Now, utilizing the positivity of  ${\hat w} ( {\bf  k} )  $ and 
$ {\hat  v} ( {\bf k}  ) $, we may apply a sine-Gordon transformation \cite{jaffe}
to the repulsive terms, $w_{jk} $, and a KHS transformation 
\cite{kac,hubbard,stra}  to the attractive terms $v_{jk} $
(where, as usual, it is most convenient to utilize periodic lattice boundary
conditions). 
Neglecting an unimportant constant factor, this yields

\begin{eqnarray}
\nonumber 
{\cal Z}  =   & &  \int \frac{
 {\cal D}  { \varphi } }{  \sqrt{  | {\bf w } |  }    } 
  \int\frac{{\cal D }  \chi }{\sqrt  { | {\bf v } | }} 
 \exp [ -  {\textstyle \frac{1}{2}} 
 \sum_{j,k} { \varphi}_j w_{jk}^{-1}  { \varphi}_k  
 -  {\textstyle  \frac{1}{2}}  \sum_{l,m}  {\chi}_l v_{lm}^{-1} {\chi}_m ] 
\\
& & 
\times 
{\rm Tr}_{ \rho  }^{ \tiny   {\cal N}} 
\left\{  z^{ {\tiny  \sum_j } {\rho}_j }
\exp [ \sum_j ( - i { \varphi}_j + {\chi}_j ) {\rho}_j ]
\right\} ,  
\end{eqnarray}
where ${\bf w } = [ \beta w_{jk} ]  = [ w_{jk}^{-1} ]^{-1}  \hspace{1em}$ 
and similarly for ${\bf v } $. 
Performing the trace over the ${\rho}_j $ then yields the transformed reduced Hamiltonian

\begin{eqnarray}
{ \cal H }  [ { \varphi}, \chi ] & =&       
 {\textstyle \frac{1}{2} }  \sum_{j,k} ( { \varphi}_j  w_{jk}^{-1} {\varphi}_k  
 +   {\chi}_j  v_{jk}^{-1} {\chi}_k   ) 
\\ \nonumber
 & & - \sum_j \ln ( 1  + z e^{ - i { \varphi}_j  + {\chi}_j  } ) .    
\end{eqnarray}

Now, in  seeking to understand the possible singularities in the reduced pressure
${\bar p} (T,z)=  ( v_0 / V )   \ln {\cal Z} $, we will, initially, 
neglect fluctuations and study the saddle point(s) which extremize the integrand  
$\exp \{ - {\cal H } [ { \varphi}, {\chi} ] \} $ in (12). 
We  expect spatially uniform solutions 
${ \varphi}_j = { \varphi}_0$,
${\chi}_j = {\chi}_0$ (all $j$) to suffice: 
from $ \partial {\cal H} / \partial { \varphi}_j = 0 \hspace{1em} $ one thus finds

\begin{equation}
\sum_k   w_{jk}^{-1} { \varphi}_0    + i \frac{  
z e^{  - i {\varphi}_0 +  {\chi}_0  }}{
1+ z e^{  - i { \varphi}_0 +  {\chi}_0   } }  =0    ,
\end{equation}
for the repulsive terms.
On premultiplying by the matrix ${\bf w} = [\beta w_{lm} ] $ 
(see, e.g., [25]) and using the Fourier
representation (9), one obtains the simpler form
\begin{equation}
 i {\varphi}_0  =   \beta {\hat w}_0  /
( 1+ z^{-1} e^{ - ( {\chi}_0 - i {\varphi}_0 ) } )      .  
\label{rep}
\end{equation}
Similarly from 
$ \partial {\cal H} / \partial { \chi}_j = 0  $ we find 
\begin{equation}
{\chi}_0   =      \beta {\hat v}_0   /
( 1+ z^{-1} e^{ - ( {\chi}_0 - i { \varphi}_0 ) } ) , 
\end{equation}
for the attractive 
terms. 

Before analyzing 
these coupled saddle-point equations, note that if one puts
\begin{equation}
{\psi} =   {\chi}_0 -i { \varphi}_0 
\end{equation}
the two equations combine simply and can be rewritten as
\begin{equation}
{\psi} e^{ - \psi}   = - z  ( \psi  + \beta U_0 ) ,
\label{rep-att}
\end{equation}
where the total strength of the pair potential 
is measured by 
\begin{equation}
U_0 \equiv {\hat U} ( {\bf 0} ) = {\hat w}_0  - {\hat v}_0 .
\end{equation}
Furthermore, by substitution of any solution of (18) on the right hand sides 
of (15) and (16) one obtains the separate solutions $ { \varphi}_0 $, which 
may evidently be imaginary, and 
$ {\chi}_0 $.

\section{TRIVIAL HARD CORES AND GAS-LIQUID CRITICALITY}
We remark, first, that following the pioneering study of Hubbard and 
Schofield \cite{hubbar2},
all authors interested in ordinary gas-liquid criticality 
have treated  the repulsive interactions in a fluid by use of a  
{\it reference system}: for a recent  example, see Brilliantov \cite{brillian}.
Unless this reference fluid is essentially trivial, as for single-site hard 
cores on a lattice, this entails increasingly detailed knowledge of the
correlation functions of the repulsive-core system \cite{hubbar2,brillian}.
Nevertheless, it will be helpful for us to
make contact with this approach by, initially, neglecting the nontrivial 
repulsive interactions embodied in $w ({ \bf r}) $. 
If the attractions are {\it also} neglected the grand partition function (11) 
reduces simply to 
$ {\cal Z} = ( 1 + z)^{\cal N} $. 
This predicts  a repulsive-core singularity at $z_0 = -1$ (all $T$) with an  exponent 
$ \phi (d=0) = 0 $(log) \cite{lai}. Evidently, the actual dimensionality, $d$, plays no role.

Now, following traditional treatments (e.g.,\cite{negele}) 
let us introduce attractive terms with 
$ {\hat v } ( {\bf k} ) > 0 $.  The KHS transformation 
then leads to the saddle-point equation (18) in which, now, 
$U_0 = -{\hat v}_0 < 0$ is negative and  we can identify 
$\psi$ directly with $ {\chi}_0$
(since $ {\hat w}_0 = 0 $ and 
$ \int {\cal D } { \varphi}$ etc., can be ignored).
Fig. 2 then provides  a graphical representation of (18) 
which can be readily analyzed: note that the bold curve depicts 
$ {\psi} e^{ - \psi } $ which has a point of inflection 
at $ {\psi}_c = 2$ and a 
corresponding  tangent that intersects the axis at 
${\psi} = 4$: see the dashed line.

By inspection, one then sees that when  $z$ increases from 
$z=0$ (along the real axis) 
for temperatures such that $ \beta { \hat v }_0 < 4 $ there 
is always a single saddle-point solution, 
$ {\psi}_0 (T, z) $, which varies analytically with $T$ and $z$: see, e.g., 
the lines labelled (a) and (b) in Fig. 2. 
Conversely, for lower $T$, when $ \beta {\hat v}_0 > 4 $, there is 
a single analytic solution, $ {\psi}_0 (T, z)$, for small $z$ 
[as on the line (c)] but
for an intermediate range of $z$ three distinct solutions arise,
as illustrated by (d); finally, for larger $z$ only a single solution remains: line (e).
Evidently the three solutions merge at a bifurcation point
determined by the inflection point at $ {\psi}_c = 2$: this leads to 
$ {\beta}_c {\hat v}_0 = 4 $ and $ z_c = e^{-2} \simeq 0.135$.
Expanding $ {\cal H} [ \chi ] $ in powers of 
$ ( \beta - {\beta}_c ) $, $(z- z_c ) $ and 
$ \delta {\chi}_j = {\chi}_j - {\chi}_c $, all taken as real variables, shows that 
this saddle-point bifurcation simply represents 
the anticipated classical or mean-field 
gas-liquid critical point at $ k_B T_c^0 = {\textstyle \frac{1}{4} } {\hat v}_0 $.
As usual, the corresponding LGW Hamiltonian can be used as a starting point 
in a field-theoretic renormalization-group (RG) treatment 
which then leads to all the standard results.

On the other hand, for $ \beta$\raisebox{-0.6ex}{ $ \stackrel{<}{\mbox 
{\scriptsize ${\sim}$}} $} 
 ${\beta}_c $, i.e., $T$\raisebox{-0.6ex} { $ \stackrel{>}{\mbox 
{\scriptsize ${\sim}$}} $} 
 $T_c $, one
can follow Ref. \cite{MEF} and discover two Yang-Lee edge singularities at 
complex $z$ (with small imaginary parts when $T$ is near $T_c$): at the saddle-point 
level these have 
$ \sigma = \frac{1}{2}$; but the fixed-point Hamiltonian 
is controlled by 
an $ i ( \delta \chi )^3 $ coupling leading, as explained above,
to an RG $\epsilon = 6-d$ expansion \cite{MEF}.

If, next, {\it small} repulsive terms, ${\hat w} >0 $, 
are introduced, the overall interaction parameter $U_0 = {\hat w}_0 - {\hat v}_0 $,
in (18) remains negative and the arguments proceed in essentially the same manner 
(although, in due course, the field $\varphi$ for the repulsive terms would 
normally be integrated out). 
As expected, neither the usual gas-liquid nor the Yang-Lee edge 
singularities undergo any change in character. 
However, if $ {\hat w}_0$   becomes sufficiently large, $U_0$ becomes 
{\it positive}  and then, clearly, the previous analysis fails!
In particular, as seen in Fig. 3 [lines (a) and (b)], for positive (real) $z$,
there is always only a single, smoothly varying saddle-point solution,
${\psi}_0 (T,z)$. 
Of course, this does {\it not} 
(necessarily) mean that the usual gas-liquid and 
Yang-Lee edge singularities are lost.
Rather, the form of the Hamiltonian, ${\cal H} [ \varphi , \chi ]$, represents 
an inadequate starting point for a perturbative RG approach:
instead, it becomes necessary to integrate out (at least to some degree) the repulsive terms, i.e.,
to perform some $\int {\cal D} \varphi $ integrals and, thereby, 
make contact with the reference-fluid treatments \cite{negele,hubbar2,brillian}.

\section{THE REPULSIVE-CORE SINGULARITY}

On  the other hand, 
when the repulsions dominate, so that $U_0 >0$ (as we will assume hereon), one sees 
from Fig. 3  [e.g., lines (c) and (d)] that 
for a real {\it negative} $z$ there is 
a unique positive saddle-point solution, say ${\psi}_{-} (T,z) < 1 $, that vanishes when 
$z \rightarrow 0 $. 
However, when $z$ approaches $ z_0 (T) < 0 $ (from above) 
a bifurcation point is reached at which 
the saddle-point 
must become complex: see the broken line in Fig. 3 that corresponds to 
$ z = z_0 (T) $. 
(At $z=z_0 $ a second, larger saddle-point solution,
${\psi}_{+}  (T,z)$, merges with ${\psi}_{-} (T,z) $. 
For $z$ much larger, additional  real {\it negative} saddle-point solutions,
${\psi}_{\pm}^{\prime}$, appear but these are not relevant 
to the dominant repulsive-core singularity.)
As evident from Fig. 3, the bifurcation point is located by the tangent passing through 
$ {\psi } = - \beta U_0 <0$: this leads to 
\begin{equation} 
0 \le {\psi}_0 (T) = {\textstyle \frac{1}{2} }  \beta U_0 
\{ [ 1 + ( 4 k_B T/ U_0 ) ]^{1/2} - 1 \} < 1 , 
\end{equation}
and 
\begin{eqnarray}
z_0 (T) & = &   - [ 1 - {\psi}_0 (T) ] \exp [ - {\psi}_0 (T) ] ,
\nonumber \\
& {\approx} & - 1 / e \beta U_0 [ 1 + {\cal O} ( 1 / \beta U_0 ) ],
\end{eqnarray}
the last result applying for strong repulsions 
($ \beta U_0 \rightarrow \infty $).
Clearly, this saddle-point bifurcation represents the repulsive-core 
singularity. 

Incidentally, if following 
Hauge and Hemmer \cite{hauge}, 
we had treated the ($d=1$)-dimensional continuum hard-rod  gas 
with additional  infinite-range, infinitely weak repulsive Kac potentials, 
we would, at this point, have found $z_0 = - 1 /e \beta U_0 $ in precise 
accord with the exact (limiting) calculations \cite{hauge}.
But, of course, our saddle-point treatment is not restricted to 
$d=1$ even though in the Kac limit it will also become exact.

To complete the analysis we may now follow standard procedures by 
expanding about the saddle point values ${\varphi}_0 $ and $ {\chi}_0 $ 
[following from (15)-(17)].
In terms of  
$\delta {\varphi}_j = {\varphi}_j  -{\varphi}_0 $,  
$\delta {\chi}_j  = {\chi}_j  -{\chi}_0 $, and 
$ \delta {\psi}_j = {\delta} {\chi}_j   - i \delta {\varphi}_j$,
the Hamiltonian truncated at fourth order becomes 
\begin{eqnarray}
{\cal H} [ \delta \varphi  , \delta \chi ] & =&  {\textstyle \frac{1}{2} }  
 \sum_{j,k}  ({\varphi}_j w_{jk}^{-1}   {\varphi}_k   
   +  {\chi}_j v_{jk}^{-1}  {\chi}_k   ) 
\\ \nonumber   
& & +  \sum_j [ {\textstyle  \frac{1}{2} }  r_0 ( \delta {\psi}_j )^2 
-g_0 ( \delta {\psi}_j )^3 
+u_0 ( \delta {\psi}_j )^4 ] ,  
\end{eqnarray}
where it is convenient to put
\begin{equation}
{\zeta}_l (T,z) \equiv -z^{-1} e^{ - {\psi}_0 } / ( 1 + z^{-1} e^{- {\psi}_0 })^l ,
\end{equation}
(which is real and positive for $z \simeq z_0 $ when 
$ l=0,2,4, \cdots $) 
so that $r_0 = {\zeta}_2 (T, z) $ and 
\begin{equation}
g_0 = {\textstyle \frac{1}{6} } {\zeta}_3 ( 1 + {\zeta}_0 ) , \hspace{2em}
u_0 = {\textstyle \frac{1}{24} } {\zeta}_4 ( 1 + 4 {\zeta}_0 + {\zeta}_0^2 ).
\end{equation}
In the usual continuum approximation this 
becomes   
\begin{eqnarray}
\nonumber
{\cal H}  =  \int d^d  { x}  [ {\textstyle \frac{1}{2}}
& & r_{11}  \delta {\varphi}^2  ( {\bf x }) + 
r_{12} \delta \varphi ( {\bf x }) \delta \chi ( {\bf x}) 
+ {\textstyle \frac{1}{2}}  r_{22}  \delta {\chi}^2  ( {\bf x}) 
\\  \nonumber
& &  +   {\textstyle \frac{1}{2}} c_{\varphi }  ( \nabla \delta \varphi  )^2 + 
 {\textstyle  \frac{1}{2} }  c_{\chi}  ( \nabla \delta \chi )^2  
\\ 
& &    -  g_0  \delta {\psi}^3  ({\bf x }) 
+ u_0 \delta {\psi }^4   ({\bf x })   ], 
\end{eqnarray}
with $c_{\varphi} = a^2 /\beta { {\hat w }_0 } $ and   
$c_{\chi} = b^2 / \beta { { \hat v} }_0  $ and 
\begin{eqnarray}
\nonumber
r_{11} & & =  {k_B T }/{ {\hat w}_0 }  -   {\zeta}_2 (z,T) , \hspace{4em}
\\ \nonumber  
 r_{22} & &  = { k_B T }/{ {\hat v}_0 }  +  {\zeta}_2 (z,T)  ,
\\ 
  r_{12} & &   =  - i    {\zeta}_2 (z,T) .
\end{eqnarray}

Before proceeding generally, let us suppose that only repulsive terms act, 
i.e., ${\hat v}_0 =0$.
Then we may drop the integrals 
$ \int {\cal D} \chi $ and see, by (17), 
that $ \delta \psi = - i  \delta \varphi$; 
but note that $ {\psi}_0 (T) $ remains real, positive and less than unity.
The continuum Hamiltonian then  reduces to 
\begin{equation}
{\cal H }_{0} [ \delta \varphi ]= \int d^d   x   
[ {\textstyle \frac{1}{2}}  t_0  ( \delta \varphi )^2 +
{ \textstyle \frac{1}{2} }  c_{ \varphi}   ( \nabla  \delta \varphi )^2   - 
i g_0   {  \delta \varphi }^3  +  u_0      {\delta \varphi }^4 ] ,
\end{equation}
where the controlling coefficient is 
\begin{equation}
t_0 (T, z)   = r_{11}  =  \frac{1}{  \beta {\hat w}_0 } 
  \left( 1 - \frac{ z }{ {z_0} }  \right)
 \frac{ [ 1- z z_0 e^{ 2 { \psi}_0 (T)  }   ] }{  [ 1 + z e^{ { \psi}_0 (T)   } ]^2 } , 
\end{equation}
which is positive for small $z$, decreases as 
$z = - |z| $ increases in magnitude, and 
vanishes linearly when $z \rightarrow z_0 (T) + <0$. 
The fourth order coefficient, $u_0 $, is positive and approaches 
\begin{equation} 
u_{00} = {\textstyle \frac{1}{4}} ( 1 -{\psi}_0 ) (1 - {\psi}_0 + 
{ \textstyle \frac{1}{6}}  {\psi}_0^2 )/
{\psi}_0^4  > 0, 
\end{equation}
when $ z \rightarrow z_0 (T) $, while 
the cubic term is purely imaginary with a coefficient 
approaching 
\begin{equation}
g_{00}  = - {\textstyle \frac{1}{6} } ( 1 - {\psi}_0 ) 
( 2 - {\psi}_0 ) /  {\psi}_0^3  < 0 .
\end{equation}

At this point we have thus reached the same stage as in the original treatment 
\cite{MEF} of the Yang-Lee edge singularities. 
The imaginary term $ i ( \delta {\varphi})^3 $ dominates the behavior. 
A momentum-shell RG analysis  reveals an upper critical dimension $d_c =6$,
an exponent 
$\eta \approx  - {\textstyle  \frac{1}{9} } \epsilon $ for   $\epsilon = 6 -d > 0 $,
and, using a unit cutoff in momentum space, a  fixed point value of 
$ g_0^* \approx ( \epsilon / 54 K_6 )^{1/2} $, 
where  $K_6$ is the area of the  unit sphere at $d=6$ dimensions.
All other results follow as previously [4,7].

Returning now to the general case of (25), 
in which attractive interactions are 
present (i.e., $ {\hat v}_0 >0 $) but $U_0 $ remains positive, 
we anticipate that only one linear combination of 
$ \delta \varphi $ and $ \delta \chi $ will become critical. 
To check this at the saddle-point level, it suffices to put 
\begin{equation}
\delta \chi ( {\bf x} ) = \delta {\chi}^{\prime} ( { \bf x} ) 
+ i ({\zeta}_2 / r_{22}  )  \delta \varphi ( {\bf x} ) .  
\end{equation}
The quadratic (nongradient) terms in (25) then become 
\begin{equation}
{\textstyle \frac{1}{2} } t ( \delta \phi )^2 + 
{\textstyle \frac{1}{2} } r_{22} (\delta {\chi}^{\prime} )^2 .
\end{equation}
The coefficient $r_{22} (T, z) $ remains positive for all 
negative $z$ but one finds 
\begin{equation}
t ( T, z)  = r_{11} - \frac{ r_{12}^2}{r_{22}}   \approx 
\frac{  t_0 (T, z)  }{[ 1 +  \beta {\hat v}_0 ( 1 - {\psi}_0  ) /{\psi}_0^2 ]} , 
\end{equation}
for $ z \simeq z_0 $: thus by (28), 
$t$ vanishes when $z \rightarrow z_0 (T)$.
Integration over the noncritical field 
$ \delta {\chi}^{\prime} ({\bf x} ) $ is 
hence justified and may be carried out perturbatively.
Apart from additive terms that are nonsingular near 
$z_0 (T) $, this finally leads to a renormalized LGW Hamiltonian 
\begin{eqnarray}
\nonumber
{\cal H }_R [ \delta \varphi ]= \int d^d  { x}  
& & [  {\textstyle \frac{1}{2}} t_R  ( \delta \varphi )^2 +
{\textstyle \frac{1}{2} }  a_R^2   ( \nabla  \delta \varphi )^2   
\\ 
& & - i h_R \delta \varphi  
- i g_R   (   \delta \varphi )^3  + \cdots  ] ,
\end{eqnarray}

where the fourth and higher order terms have been dropped. 
For ${\hat v}_0 $ sufficiently small, the renormalized 
couplings $t_R $ and $g_R$ will differ little from 
$t$ and $g$ and, in particular, 
$h_R$ and $g_R$ will be real, so that the $ i ( \delta \varphi )^3 $ 
term once again dominates.

Stability of the saddle point requires that the 
renormalized repulsive range, 
$a_R$, in (34)  be real (and positive). 
Roughly, we expect $a_R^2  \propto (a^2 {\hat w}_0 - b^2 {\hat v}_0 )$: 
the positivity of this factor then represents a previously 
unstated restriction on the initial potentials.
In practice,
however, if this (or the more precise) condition fails, 
it may be sufficient, as explained in the discussion of gas-liquid 
criticality, to perform further, nonperturbative renormalizations 
to dampen  the attractive interactions for $z<0$ and so expose again a saddle-point 
representation of the repulsive-core 
singularity at $z_0 (T) $. Certainly, we must expect the form (34) to 
represent the singularity whenever it is actually realized in a system; and, then, 
clearly it must belong to the Yang-Lee edge universality class.

\section{CONCLUSIONS}
In summary,  by using separate field-theoretic transformations for the 
repulsive and attractive parts of the pair interactions in a fluid, 
we have demonstrated in general the presence of a universal repulsive-core
singularity on the negative axis at a value 
$ z_0 (T) <0 $: see Fig. 1. The behavior of the pressure in the vicinity of 
$z_0$ is stated in (6). The singularity belongs to the same 
universality class as Yang-Lee edge criticality, 
as proposed by Lai and Fisher \cite{lai}: see Eqs. (1)-(4) and Fig. 1. 
The basic critical exponents are related via (7) (see also \cite{lai});
the borderline dimensionality is $d_c = 6$; and an $i {\varphi}^3 $ coupling characterizes 
the LGW fixed-point Hamiltonian \cite{MEF}.

%
%

\section*{Acknowledgments}
Y. P thanks  the Institute for  
Physical Science and Technology at the University of Maryland for 
hospitality. Her work  has been supported in part by the Ministry of 
Education of Korea  through the Basic
Science Research Institute of Seoul National University.
M.E.F. acknowledges support from the U. S. National Science 
Foundation (through Grant CHE 96-14495). 
The interest of Youngchan Kim has been
appreciated.

 



\begin{figure}
\hskip -0.4cm
\epsfxsize=8cm \epsfysize=6cm \epsfbox{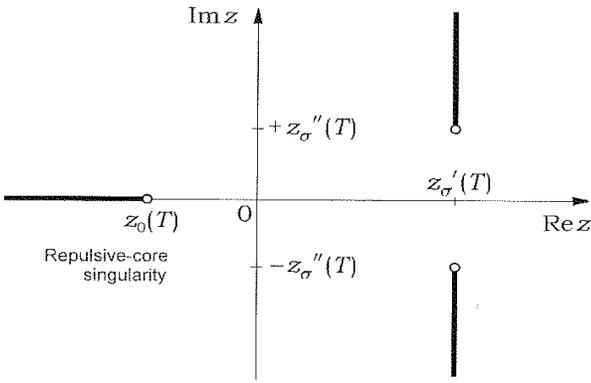}
\narrowtext

\caption{ 
Complex activity plane for a  lattice or continuum fluid system 
at temperatures $T > T_c$, above gas-liquid criticality.
The Yang-Lee edge singularities are   located at 
$z =  z_{\sigma}^{\prime } (T) \pm  i  z_{\sigma}^{\prime \prime } (T)  $ 
while  the repulsive-core singularity 
lies on the real negative activity axis $z=z_0 (T) $. 
(Note that the branch cuts running from the Yang-Lee edges will {\it not},
in  general, be linear as shown,  purely schematically, here: indeed, for a simple lattice-gas the 
cuts lie on circles centered at $z=0$.) } 
\label{fig:phase}
\end{figure}
\begin{figure}
\epsfxsize=8cm \epsfysize=6cm \epsfbox{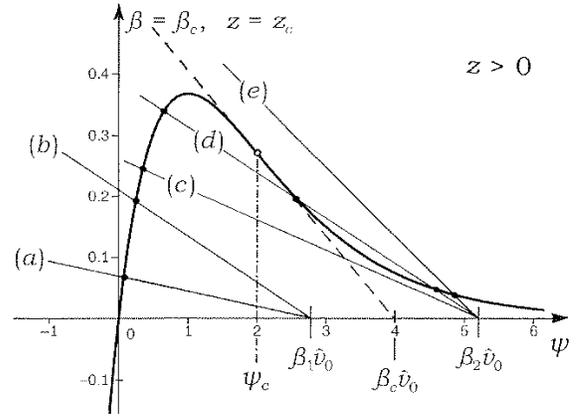}
\caption{
Plot of ${\psi} e^{ - \psi }$ (bold curve) versus $ \psi$ to elucidate 
solutions of the saddle-point equation (18) 
\hspace{0.01em}   in \hspace{0.01em}
 the  case   of  \hspace{0.01em} only \hspace{0.01em}
single-site  \hspace{0.01em} repulsive      \hspace{0.01em} hardcores   
 \hspace{0.01em}     
with  \hspace{0.01em} attractive 
 \hspace{0.1em}  potentials   of strength  
$  {\hat v}_0 $ ($ =  - U_0 $) for various temperatures,
$ {\beta}_1$, ${\beta}_c$ and $ {\beta}_2$ and  positive activities $z$.
The open circle marks the point of inflection which serves to 
locate the (classical) gas-liquid critical point; the dashed line is the associated 
tangent. The solution lines have slopes $ -z$ in accord with (18).}

\end{figure}
\begin{figure}
\epsfxsize=8cm \epsfysize=6cm \epsfbox{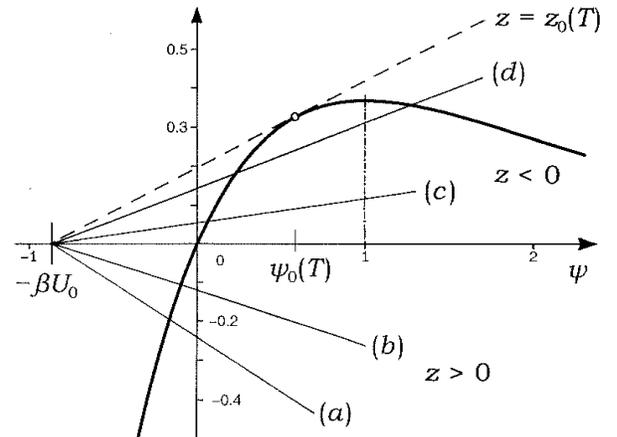}
\caption{
Plot of  ${\psi} e^{ - \psi }$ to illuminate the saddle-point equation (18) as in Fig.2;
but note the change in scales.
The dominant repulsive case with  
$U_0 = {\hat w}_0 -  {\hat v}_0 > 0 $ is illustrated for positive activities $z$ 
[lines (a) and (b)] and negative activities: (c) and (d) and the tangent (dashed line)
which locates the repulsive core-singularity at 
$z=z_0 (T) <0$ with field ${\psi}_0 (T) $. }
\end{figure}


\end{multicols}
\end{document}